\documentclass[12pt,a4paper]{article}
\usepackage{times}

\usepackage{lscape}
\usepackage{graphicx,epsfig}
\usepackage[tbtags]{amsmath}

\begin{document}


\title{%
Large Order Asymptotics and Convergent Perturbation Theory for
Critical Indices of the $\boldsymbol{\phi ^4}$ Model in
$\boldsymbol{4-\epsilon }$ Expansion. }

\author{
J.~ Honkonen\,\thanks{E-mail address: Juha.Honkonen@helsinki.fi}\ \ $^1$, M.~
Komarova\,\thanks{E-mail address: komarova@paloma.spbu.ru}\ \ $^2$, M.~
Nalimov\,\thanks{E-mail address: Mikhail.Nalimov@pobox.spbu.ru}\ \ $^2$ \\
$^1$Department of Physical Sciences,
University of Helsinki,\\ and National Defence College, Helsinki, Finland\\
$^2$Department of Theoretical Physics, St.~Petersburg State
University,\\ St.~Petersburg, Russia }

\date{}

\maketitle

\abstract{%
Large order asymptotic behaviour of renormalization constants in
the minimal subtraction scheme for the $\phi ^4$ $(4-\epsilon)$
theory is discussed. Well-known results of the asymptotic
$4-\epsilon $ expansion of critical indices are shown to be far
from the large order asymptotic value. A {\em convergent} series
for the model $\phi ^4$ $(4-\epsilon)$ is then considered. Radius
of convergence of the series for Green functions and for
renormalisation group functions is studied. The results of the
convergent expansion of critical indices in the $4-\epsilon $
scheme are revalued using the knowledge of large order
asymptotics. Specific features of this procedure are discussed.}


\section{Introduction}
\label{sec:intr} \setcounter{section}{1}\setcounter{equation}{0}
 Calculation of critical indices is usually based on
a certain asymptotic expansion. To obtain reliable results a
resummation procedure is necessary. To this end the Borel-Leroy
transform in fixed dimension \cite{fixedD,Guida} or in $\epsilon$
expansion \cite{Guida,epsBo,Guillou}, the simple Pad\'e-Borel
method \cite{Antonenko} as well as self-similar exponential
approximants \cite{others} have been used. But divergent series
can produce an arbitrary result. Thus some additional information
about the series is needed.

When a convergent series is considered, additional
information can also improve the result. The knowledge about
location and character of singularity of the function investigated or
about the large order asymptotic behaviour of the expansion may be used
to accelerate convergence of the series. In this report we will discuss
a resummation of the $4-\epsilon $ expansion of critical indices for
the $\phi ^{4}$ theory with help of the large order asymptotics.

\section{Large order asymptotics of renormalization constants} \label{sec-1}

Large order asymptotic behaviour of the  $O(n)$ symmetric $\phi ^4$  theory is well
known. The N-th order term of the standard expansion for an arbitrary
function $F$ in the theory with the renormalized action
\begin{equation}
S_{R}(\phi ,g)={1\over 2}Z_{\phi }^2\partial _i{\phi
_\alpha}\partial _i{\phi _\alpha}+{1\over 2}Z_{\phi }^2Z_{\tau
}\tau {\bf \phi }^2+{1\over 4!}Z_{\phi }^4Z_gg\mu^{\epsilon }({\bf
\phi }^2)^2 \label{act}
\end{equation}
behaves at large $N$ as
(\cite{Lipatov, ZinnBr, Zinn})
\begin{equation}
F^{(N)}\approx N^Ne^{-N}a^NN^{b_F}c_F. \label{1}
\end{equation}
The notation $F^{(N)}$ for the $N$-th order coefficient of the
expansion of the function $F$  will be used henceforth. Working
within the minimal subtraction (MS) scheme in the $4-\epsilon $
expansion we have corrected the result (\ref{1}). Namely,
analysing the limits $N\to \infty $ and $\epsilon \to 0$ in more
detail, we have obtained a more accurate estimate for the
amplitudes $c_F$ of the renormalization constants \cite{TMF}.
According to the results of Ref. \cite{Lipatov} the asymptotic
expression (\ref{1}) for the $\beta $ function coincides
practically with the exact result for $N=3$, but the asymptotics
of \cite{Lipatov} for $N=4, 5$ are larger than the exact value.
Contrary to these conclusions, we have shown that the asymptotics
(\ref{1}) for the renormalization constants (and for the critical
indices) are much smaller than the exact ones \cite{Tck}. For
example, $[Z_{g}^{(5)}]_{asymp}\approx .01[Z_g^{(5)}]$ for $n=1$,
where $[ Z_{i}]$ is the residue of $Z_i$ regarded a function of
complex $\epsilon$. We have predicted that only starting from 10 -
15-th order the perturbation expansion could be near to the
asymptotic value. Due to this fact we can state that the Borel
transform of the 5 known terms of the $\epsilon$ expansion has no
theoretical ground.

Nevertheless, we propose extrapolation expressions for the unknown
terms in the expansions of renormalization constants $Z$ in the following form
\cite{TMF}:
$$[\bar Z_g^{(N)}]_{asymp}= [Z_{g}^{(N)}]_{asymp}\Bigg
( 1 + \frac{\bar c_g}{N}\Bigg ), \quad \bar
c_g=\frac{5[Z_{g}^{(5)}]}{[Z_{g}^{(5)}]_{asymp} } - 5$$
\begin{eqnarray}
\label{extr} [\bar Z_{\phi }^{(N)}]_{asymp}= [Z_{\phi
}^{(N)}]_{asymp} \Bigg (1 +\frac{\bar c_{\phi}}{N} \Bigg),\quad
\bar c_{\phi}=\frac{5[Z_{\phi }^{(5)}]}{ [Z_{\phi
}^{(5)}]_{asymp}}  - 5.
\end{eqnarray}
These expressions contain an additional correction of the $1/N$
type normalized by direct comparison of the asymptotic
value (\ref{1}) for $N=5$ with the exact one. In (\ref{extr}),
\begin{eqnarray}
\label{Zg} [Z_{g}^{(N)}]_{asymp}=(-1)^N C_gN^N
N^{\frac{7+n}{2}}e^{-N}, \quad [Z_{\phi }^{(N)}]_{asymp} =(-1)^N
C_{\phi} N^N N^{\frac{3+n}{2}}e^{-N}
\end{eqnarray}
\begin{eqnarray}
\label{tri} \nonumber
C_g=-\frac{864\pi^{2+n/2}D_{et}}{n(n+2)\Gamma(n/2)}\exp {\Bigg
(}\frac {n+8}{6}{\Bigg [}\Psi(1)-\ln (\pi )-2{\Bigg ]}{\Bigg )}, \\
C_{\phi}= -\frac{6\pi^{2+n/2}D_{et}}{n\Gamma(n/2)}\exp {\Bigg
(}\frac {n+8}{6}{\Bigg [}\Psi(1)-\ln (\pi )-2{\Bigg ]}{\Bigg )},
\end{eqnarray}
$$D_{et}=2^{(7-4n)/2}3^{5n/2}\pi
^{-(5+n)/2}5^{-5/2}e^{-4(n+2)/3 - 7(n+8)/18 - R/12+
(n+8)(\Psi(1)+2-ln(\pi))/6},$$
\begin{eqnarray}
\label{jun}
R:=-\sum_{l=2}^{\infty}\sum_{p=3}^{\infty}\frac{(l+1)(l+2)(2l+3)}{p(l+1)^p(l+2)^p}
\Bigg (6^p+2^p(n-1) \Bigg )
\end{eqnarray}
Here, $\Psi$ is the logarithmic derivative of the $\Gamma$
function.
Using similar expressions normalized with help of the 4-th exact
term of the renormalization constant expansions, we could find the 5-th
order term and estimate the accuracy of such calculation as at
least 80\%. Besides, we hope that the expressions
(\ref{extr}) can give us a
reliable estimate for more than 5 orders of the expansions of $Z$'s.

\section{Convergent expansion for critical exponents} \label{sec-2}

We will use the knowledge of large order asymptotics to improve
initially convergent series for critical exponents. In Ref. \cite{Juha}
a new approach was suggested to calculate critical indices. It was
based on the modified expansion for the $O(n)$-symmetric
$\phi^4$ model \cite{Yshver} which leads to {\em convergent}
series. The standard action $S=S_1+S_2$ with $S_1=\int dx(\partial
\phi )^2/2$, $S_2=g\int dx \phi ^4/4!$ was cast in the form
$S=S_0+S_I$, where $S_0= S_1+aS_1^2$ and $S_I=\zeta (S_2-aS_1^2)$.
Here, $a$ is an arbitrary constant and $\zeta $ is a new expansion
parameter. The model coincides with the initial one at the
'physical' value of $\zeta $: $\zeta _{ph}=1.$ It was stated in
Ref. \cite{Yshver} that the $\zeta $ expansion is convergent for $\zeta
\le \zeta _c=1$, when
\begin{equation}
\label{amin} a\ge \frac {g}{64\pi ^2}=a_{min}.
\end{equation}
In the framework of the $\zeta $ expansion the renormalized $2k$ point
Green function
can be written as an integral with respect to an additional
variable $\sigma $ as:
\begin{eqnarray} G_{2k}^R&=& \frac 1{\sqrt {\pi
}}\int\! d\sigma e^{- \sigma ^2}\! \int\! D\phi\, \frac {\phi
({\bf x}_1)...\phi ({\bf x}_{2k})} {(1+2i\sigma \sqrt {a(1-\zeta
)})^k }\tilde Z_{\phi }^{-2k} \nonumber
\\
&\times&\exp \left(-\frac 12 \partial \phi \partial \phi  -\frac
{\zeta g\tilde Z_g\mu ^{\epsilon }\phi ^4}{4!(1+2i\sigma \sqrt
{a(1-\zeta )})^2}\right)\,. \label{2}
\end{eqnarray}
Here, $g$ is the renormalized charge, $\mu $ is the renormalization
mass, $\tilde Z_i=Z_i(g\zeta/(1+2i\sigma \sqrt {a(1-\zeta )}\hskip
0.1cm)^2)$, and $Z_i(g)$ are the renormalization constants in the
usual perturbation theory. This representation allows to construct
the Feynman graphs in the usual manner. Due to the integration with
respect to $\sigma$ every diagram acquires an additional factor
\cite{Yshver}:
\begin{eqnarray}
\label{UN} U_j(a)&=&1/\sqrt {\pi }\int d\sigma \exp(-\sigma
^2)/(1+2i\sigma \sqrt {a})^{j}
\nonumber\\
&=& \int _0^{\infty}dtt^{j-1}exp(-t-at^2)/(j-1)!
\end{eqnarray}
The basic renormalisation group (RG) equation  for the
Green functions (\ref{2}) was derived in Ref. \cite{Juha}
in the form
\begin{equation}
 {\cal D}_{RG}G_{2k}^R\equiv [\mu \partial _{\mu }+\beta \partial _g+\beta _{2}
\partial _g^2+...+\gamma]G_{2k}^R=0,
\label{RG}
\end{equation}
 where $\gamma $ and all $\beta $ functions depend on the
parameters $g$, $\zeta $, $a$ and $k$. Using the MS
scheme one can write for the RG functions in $D=4-\epsilon $
dimensional space ($n=1$, $k=1$) the expressions
\begin{equation}
\label{gamma} \gamma =-\frac {2}{U_1(a(1-\zeta))} g\partial_g \sum
_{l=0}g^l\zeta ^l [Z_{\phi }^{(l)}]U_{2l+1}(a(1-\zeta)),
\end{equation}
\begin{eqnarray}
\label{beta} \beta &=&-\epsilon g +\frac 1{U_3(a(1-\zeta))}
g\partial_g \sum _{l=0}g^l\zeta ^l \bigg([Z_{g }^{(l)}]-2[Z_{\phi
}^{(l)}]\bigg)U_{2l+3}(a(1-\zeta))-\gamma g \nonumber
\\
&\equiv& -\epsilon
g+\overline{\beta}(g)g
\end{eqnarray}
It was shown that Eq. (\ref{RG}) governs the large-scale asymptotic
behavior of the model. Critical exponents are related to the
anomalous dimensions in the usual way \cite{Zinn}: $ \eta = \gamma
(g_*)/k $, and $ 1/\nu = 2-\gamma _{\tau }(g_*)+k\eta $ [$\,\gamma
_{\tau }$ is determined by equation (\ref{RG}) for the Green
functions with the insertion of the composite operator $\phi^2\,$].
Here $g_*$ is the fixed point determined by the usual equation $\beta
(g_*)=0$, or
\begin{equation}
\label{fpC} \epsilon =\bar \beta (g_*).
\end{equation}
In Ref. \cite{Juha} it was demonstrated that it is possible to solve
Eq. (\ref{fpC}) iteratively to calculate $g_*$ in the form
of a double expansion in $\epsilon $ and $\zeta $. As a result, the
exponents $\eta $ and $\nu $ were calculated for $n=1$, $k=1$,
$a=0.134$ at the physical values $\zeta =\epsilon =1$  (see the
table). The columns correspond to expansion order taken into
account. Exponents marked by $\epsilon $ are the results of the
usual $\epsilon $ expansion, which are quoted for comparison.

\bigskip
\begin{tabular}{|l|c|c|c|c|c|}
\hline
Exponent&1&2&3&4&5\\
\hline
$\eta $&0&0.02553&0.04342&0.039745&0.039741\\
\hline
$\eta _{\epsilon }$&0&0.01852&0.03721&0.02888&0.05454\\
\hline
$\nu $&0.621&0.671&0.663&0.674&0.651\\
\hline
$\nu _{\epsilon }$&0.583&0.627&0.607&0.678&0.461\\
\hline
\end{tabular}
\bigskip

Estimates based on the Borel transform of $\epsilon$
expansion are $\eta = 0.0360\pm 0.0050$ \cite{Guida},
$\eta=0.035\pm 0.002$ \cite{epsBo}, $\nu = 0.6290\pm 0.0025$
\cite{Guida}, $\nu=0.628\pm 0.001$ \cite{epsBo}. A typical
lattice result is $\nu = 0.6305\pm 0.0015$ \cite{Zinn}. Thus, in
spite of convergence of the $\zeta $ expansions, the accuracy of the
results of \cite{Juha} is lower than in case
of estimates based on Borel transforms. This is why we will
improve the convergent
expansion using the large order asymptotics.

\section{Large order asymptotics for the convergent perturbation expansion}
\label{sec-3}

In Ref. \cite{Yshver} it was stated that instanton analysis
could not be used for the investigation of the convergent $\zeta
$-expansion of Green functions. Using more sophisticated
examination, the radius of convergence was estimated as $\zeta
_c=1$.

Contrary to this, we confirm the adequacy of instanton approach here.
Indeed, let us determine the $N$-th order of $\zeta $ expansion by
$$
G_{2k}^{(N)}=\oint \frac {d\zeta}{\zeta ^{N+1}}G_{2k},
$$
where $G_{2k}$ is given by (\ref{2}). A subsequent steepest descent
approach in the variables $\sigma $, $\zeta $ and $\phi $  leads
to an instanton and stationary point (for $n=1$)
$$\phi _s =\sqrt N\phi _0\Bigg(\frac {2\sigma _0\sqrt {a(1-\zeta _s)}}{\sqrt{
\zeta _sg}}-\frac {i}{6\sqrt{\zeta _s gN}}+O(\frac 1N)\Bigg),
$$
where
$$
\phi _0=\frac {4\sqrt 3}{ y}\frac 1{1+({\bf x}-{\bf x}_0)^2/y^2},
$$
with arbitrary ${\bf x}_0$, $y$, and
$$\zeta _s = \bar \zeta _c\Bigg(1+\frac{7}{6\sqrt {aN}}+O(\frac
1N)\Bigg), \hskip 1cm
\bar \zeta _c=\frac 1{1-g/(64\pi^2a)}
$$
$$\sigma _s=\sqrt N\sigma _0,
\hskip 1cm \sigma _0=\sqrt {\bar \zeta _c} \sqrt {\frac {g}{64\pi
^2a}}+\frac {\bar \zeta _c}{\sqrt N}\frac {568a\pi ^2+3g}{96\pi
a\sqrt {g}} +O(\frac 1N).
$$
The instanton analysis results in the following behavior for the
$N$-th order of the $\zeta $-expansion of the $2k$ point Green function:
\begin{equation}
\label{as} G_{2k}^{(N)}\sim N^\alpha \bar\zeta_c^Ne^{-\sqrt
 {\frac Na}}
\end{equation}
with some $\alpha$.

Contrary to \cite{Yshver} we have obtained a radius of
convergence $\bar\zeta _c$ which tends to infinity as $a\to
a_{min}$. We consider the unrenormalized Green function in $D=4$
dimensional space only, but this is not essential for the
treatment of the radius of convergence.
The same result (\ref{as}) can be obtained directly by the method
proposed in \cite{Yshver}. In such a case the instanton $\phi _c\sim
N^{1/4}$. Thus, the instanton does exist.

For usual divergent series of $\epsilon $ expansions the large
order asymptotics of Green functions characterise unambiguously
the asymptotics of RG functions. In the case of convergent series
the situation is more difficult. Namely, the denominators of the
expressions (\ref{gamma}), (\ref{beta}): $U_i(a(1-\zeta ))$
($i=1,3$) have the radius of convergence equal to 1. Therefore,
the large order asymptotics of the $\zeta $ expansion of these
functions
$$U_i^{(N)}(a(1-\zeta ))\approx N^{i/2-1}e^{-\sqrt {N/a}+1/(8a)}a^{1-i}/
\Gamma (i)$$
have a non-trivial influence on the asymptotic behaviour of
$\beta$ and $\gamma$ functions.

\bigskip
\section{Character of singularity of the convergent expansion for critical indices}
\label{sec-4}

In addition to $\zeta _c$ problem we have some difficulties with
double $\zeta $, $\epsilon $  expansion of indices as we try to
find the fixed point $g_*$ of (\ref{fpC}). The iteration solution
$g_*(\epsilon, \zeta)$ of the equation (\ref{fpC}) has a new
singularity at the point $\epsilon _m$ bounded by the nearest to
zero extremum of the $\bar \beta $ in the complex plane $g$.
Namely, the point $g_m$ given by
\begin{equation}
\label{sing}
\epsilon _m=\bar \beta (g_m, \zeta),
\hskip 1cm
\partial _g \bar \beta (g_m)=0
\end{equation}
results in the singularity $g_*\sim \sqrt {\epsilon -\epsilon
_m}$.

To calculate $\epsilon _m$ the second equation (\ref{sing}) has to
be solved in a form of a $\zeta $ expansion. Unfortunately, our
calculation shows that the five known terms of the
expansion of $\bar \beta$ are not sufficient for a reliable
determination of $g_m$.
If the extrapolation expressions (\ref{extr}) are used
within the Pad\'e
approximant approach, then
there is a whole set of conjugate
points $g_m$. It is very difficult to find a suitable 'mapping'
to exclude these singularities. Moreover, the corresponding minimal
value of $\epsilon _m$ turns out to be very small ($\sim
.02$) that decreases the radius of convergence in $\epsilon$ of
our double expansion. Thus the method to be used is to consider
the  variables $g$, $\zeta $ instead of $\epsilon $, $\zeta $. A similar
approach leads to a good result in the Kraichnan model, where
a convergent series is dealt with as well \cite{Adzhemyan}.

Calculating $g_*$ directly from the equation (\ref{fpC}) for
$\epsilon =1$ we were able to obtain five orders of the $\zeta $
expansion for critical indices (for $n=1$, $\zeta _{ph}=1$):

\bigskip
\begin{tabular}{|l|c|c|c|c|c|}
\hline
Exponent&1&2&3&4&5\\
\hline
$\eta $&0&.021 &0.023 &0.027 &0.029 \\
\hline
$g_*/(16\pi^2)$&0.96 &0.86 &0.65 &0.61 & 0.57\\
\hline
$\nu $&0.612&0.650&0.638&0.641&0.640\\
\hline
$g_*/(16\pi^2)$&1.38&1.17&0.84&0.75&0.68\\
\hline
\end{tabular}
\bigskip

The exponent $\eta $ was calculated for the value
$a=.14 $, the exponent $\nu $ for $a=.17$ which lead to the best rate of series
convergence. Note that $g_*$ obtained ensures the convergence of
the $\zeta $ series in both cases. However, using the extrapolation
expressions (\ref{extr}) for unknown terms of the expansions of the
renormalisation constants $Z$, we
obtained the results shown in Fig. 1  demonstrating
the failure of the usual extrapolation procedure in the calculation
of critical indices.

\begin{figure}[t]
\epsfig{file=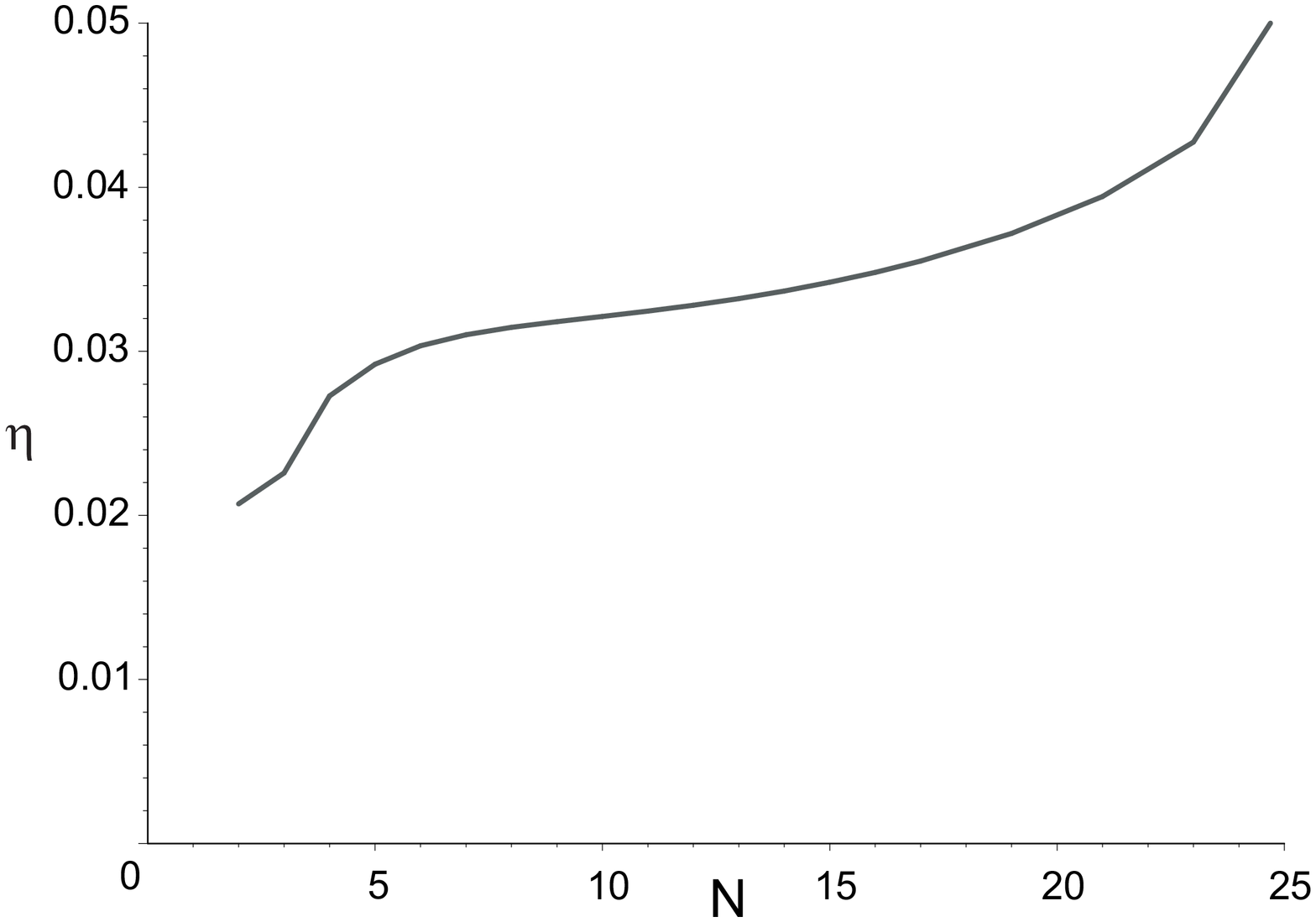,width=8 cm, angle=270}\\
\\
\epsfig{file=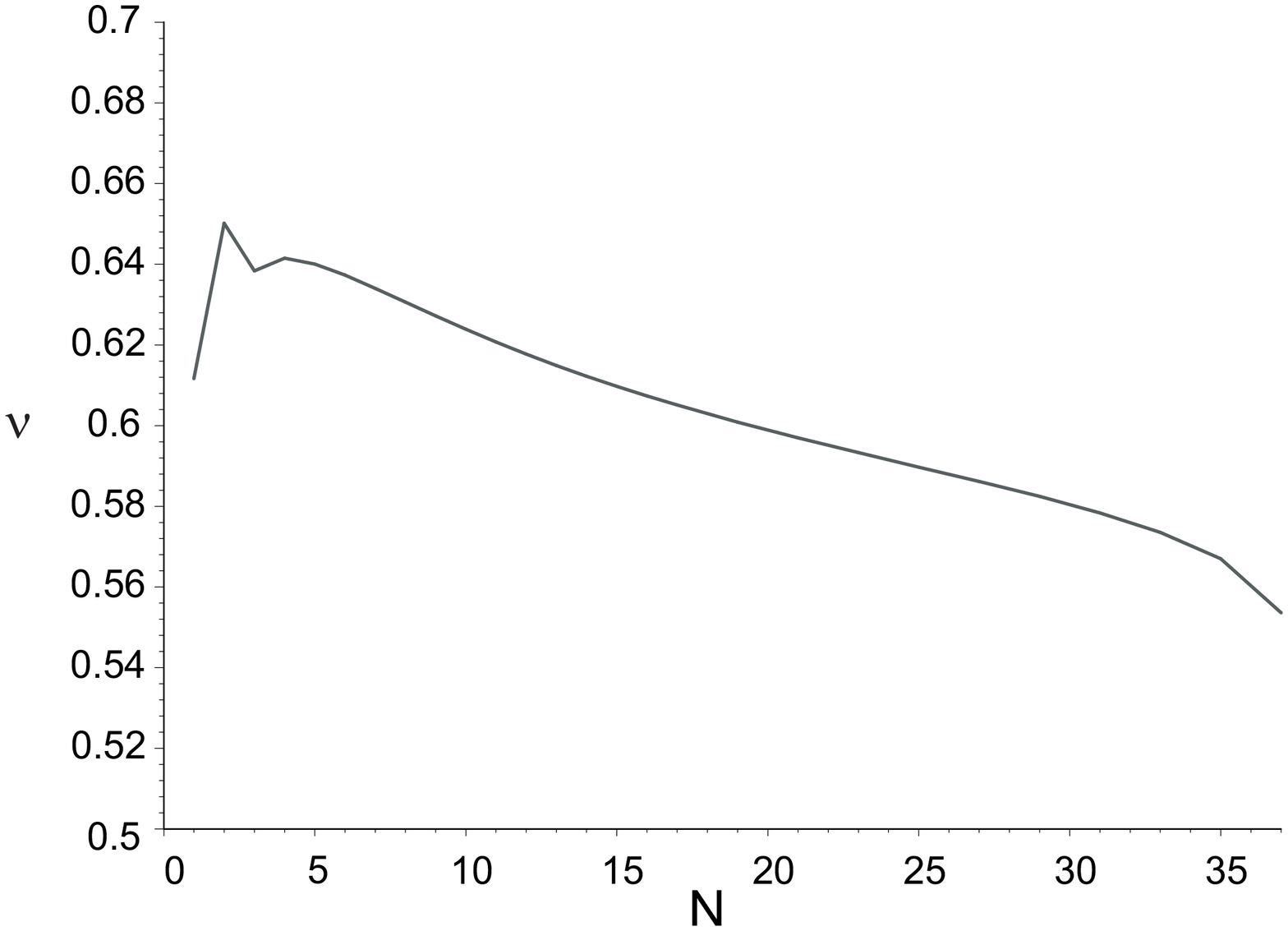,width=8 cm, angle=270} \\
\\   \caption{The value of the index $\eta$ and
the index $\nu$  as a function of perturbation order number $N$
taken into account.}
\end{figure}

To improve the convergence of the $\zeta $ series we can
investigate the large order asymptotics of the functions
(\ref{gamma}), (\ref{beta}). Denominators of these expressions
have essential singularities of the type $U_{1}(a(1-\zeta)$,
$U_{3}(a(1-\zeta)$ that must be extracted. It is convenient to
introduce $\delta Z_{\phi }^{(N)}\equiv [Z_{\phi }^{(N)}]-[\bar
Z_{\phi }^{(N)}]_{asymp}$. Let us rewrite the asymptotic
expression (\ref{Zg}) of the renormalization constant $Z_{\phi }$
in the equivalent form
$$[\bar
Z_{\phi }^{(N)}]_{asymp}= C_{\phi }(-1)^N N^2\frac{(2N)!}{4^N\sqrt
{2}N!}(1+\frac{c_{\phi}}{N}).$$
Here, the constant $c_{\phi}$ can
be found in a way similar to $\bar c_{\phi}$ in (\ref{extr}) by comparison
of $[\bar Z_{\phi}^{(5)}]_{asymp}$ with the exact expression $[Z_{\phi}^{(5)}]$.

Then substituting $[\bar Z_{\phi}^{(5)}]_{asymp}$ into
(\ref{gamma}) and re-expanding it in $\zeta$ one can write $\gamma
$ in the form
\begin{eqnarray}
\label{gs} \gamma =-2\sum_{j=1}^5 \frac{j(g\zeta)^j}{(16\pi^2)^j}
\frac{ U_{2j+1}(a(1-\zeta))\delta
Z_{\phi}^{(j)}}{U_1(a(1-\zeta))}- \frac{\sqrt {2}C_{\phi
}}{U_1(a(1-\zeta))}\times\ \nonumber
\\
\times\Bigg(-\frac {(g\zeta)^3}{(64\pi ^2)^3}6!U_7+\frac
{(g\zeta)^2}{(64\pi ^2)^2} 4!(3+c_{\phi })U_5 -\frac
{(g\zeta)}{(64\pi ^2)}2!(1+c_{\phi })U_3\Bigg)\,.
\end{eqnarray}
Here, the omitted for brevity argument of the $U_i$ functions is $a(1-\zeta)+
{g\zeta}/{64\pi ^2}$.
Thus, we have investigated not only the location of $\zeta _c$ and
$\bar \zeta _c$, but also the character of the singularities of
the $\zeta $ expansion of the $\gamma $ function.

In analogy with (\ref{gs}) we obtain for the $\beta $ function
\begin{eqnarray}
\label{bs} \beta =\sum_{j=1}^5 j\frac{(g\zeta)^j}{(16\pi^2)^j}
\frac {U_{2j+3}(a(1-\zeta))(\delta Z_g^{(j)}-2\delta
Z_\phi^{(j)})}{U_3(a(1-\zeta))}+ \frac {C_{g
}}{2^{5/2}U_3(a(1-\zeta))}\times \ \nonumber
\\ \times\Bigg(-\frac {(g\zeta)^3}{(64\pi ^2)^3}8!U_9+\frac
{(g\zeta)^2}{(64\pi ^2)^2}6!(3+c_{g})U_7-\frac {(g\zeta)}{(64\pi
^2)}4!(1+c_{g })U_5\Bigg)-\gamma
\end{eqnarray}
The expressions (\ref{gs}), (\ref{bs}) are physically meaningful at $\zeta=1$.
Solving numerically Eq. (\ref{fpC}) with $\beta $ from
(\ref{bs}) and substituting the value $g_*=0.382$ obtained in $\gamma $
(\ref{gs}) we obtain $\eta =0.0236$. Taking into account the
$\epsilon $ expansion up to fourth order one obtains similarly
$\eta _4=0.0241$.

An analogous procedure for the index $\nu $ leads to $\nu =0.580$,
$(\nu _4 =0.624).$ These results are the best we can obtain from
the $\epsilon $ expansion in the convergent scheme using all
available information about the large order behaviour.

It is worthwhile noting that the accuracy of our results is
similar to the accuracy of the  approximation (\ref{extr}) determined
by the rate with which asymptotics of renormalization constants tend to
the exact value. Thus, we conclude that the accuracy of the
$\epsilon$ expansion resummation with help of the Borel transformation
is lower than commonly quoted.

{\bf Acknowledgement:}
The authors are grateful to the Academy of Finland (Grant No.
75117), Nordic Grant for Network Cooperation with the Baltic
Countries and Northwest Russia No. FIN-18/2001, and in part to the
Russian State Committee for Higher Education (Grant  No.
E00-3.1-24) and Scientific Program (Russian Universities) for
financial support.


\begin{thebibliography}{99}
\bibitem{fixedD} G. A. Baker, Jr., B. G. Nickel, D. I. Meiron: {\sl Phys. Rev. B}
{\bf 17} (1978)  1365; J. C. Le Guillou, J. Zinn-Justin: {\sl
Phys. Rev. Lett.} {\bf 39}
 (1977) 95; {\sl Phys. Rev. B} {\bf 21} (1980) 3976.
\bibitem{Guida}
R. Guida, J. Zinn-Justin: {\sl J. Phys. A} {\bf 31} (1998) 8103.
\bibitem{epsBo}
S. G. Gorishny, S. A. Larin, F. V. Tkachev: {\sl Phys. Lett. A}
{\bf 101} (1984) 120.
\bibitem{Guillou}
J. C. Le Guillou, J. Zinn-Justin: {\sl J. Phys. Lett. (Paris)}
{\bf 46}, (1985) L137; {\sl J. Phys. (Paris)} {\bf 48} (1987) 19;
{\sl J. Phys. (Paris)} {\bf 50} (1989) 1365
\bibitem{Antonenko}
S. A. Antonenko, A. I. Sokolov: {\sl Phys. Rev. E} {\bf 51} (1995)
1894.
\bibitem{others} V. I. Yukalov, S. Gluzman: {\sl Phys. Rev. E} {\bf 58},
(1998) 1359.
\bibitem{Lipatov} L. N. Lipatov: {\sl Zh. Eksp. Teor. Fiz.} {\bf 72} (1977) 411.
\bibitem{ZinnBr} E.Brezin, J. C. Le Guillou, J. Zinn-Justin:
{\sl Phys. Rev. D} {\bf 15} (1977) 1544.
\bibitem{Zinn} J.  Zinn-Justin: {\sl Quantum Field Theory and
Critical Phenomena} (Oxford Univ. Press, Oxford, 1989).
\bibitem{TMF} M.V.Komarova, M.Yu.Nalimov: {\sl Theor. and Math. Phys} {\bf 126},
(2001) 393.
\bibitem{Tck} H.  Kleinert, J. Neu, V. Schulte-Frohlinde, K. G.
Chetyrkin, S. A.  Larin: {\sl Phys.  Lett. B} {\bf 272} (1991) 39;
{\sl Phys. Lett. B} {\bf 319} (1993) 545.
\bibitem{Juha} J.Honkonen, M.Nalimov: {\sl Phys. Let.} {\bf B 459}
(1999) 582.
\bibitem{Yshver} A. G. Ushveridze: {\sl Yad. Fiz.} {\bf 38} (1983) 798.
\bibitem{Adzhemyan}  L.Ts.Adzhemyan private communications.
\end{thebibliography}
\end{document}